# TSDS: high-performance merge, subset, and filter software for time series-like data

R.S. Weigel, D. M. Lindholm, A. Wilson, and J. Faden

# 1. Abstract

Time Series Data Server (TSDS) is a software package for implementing a server that provides fast super-setting, sub-setting, filtering, and uniform gridding of time series-like data. TSDS was developed to respond quickly to requests for long time spans of data. Data may be served from a fast database, typically created by aggregating granules (e.g., data files) from a remote data source and storing them in a local cache that is optimized for serving time series. The system was designed specifically for time series data, and is optimized for requests where the longest dimension of the requested data structure is time. Scalar, vector, and spectrogram time series types are supported. The user can interact with the server by requesting a time series, a date range, and an optional filter to apply to the data. Available filters include strides, block average/minimum/maximum, exclude, and inequality. Constraint expressions are supported, which allow such operations as a request for data from one time series when a different time series satisfied a specified relationship. TSDS builds upon DAP (Data Access Protocol), NcML (netCDF Mark-up language) and related software libraries. In this work, we describe the current design of this server, as well as planned features and potential implementation strategies.

# 2. Introduction

In recent years, significant progress has been made in the area of data discovery and search in the domain of heliophysics [King et al., 2008; Narock and King, 2008]. The result of a search for data is often a list of links to data files ("granules"). Often these files are large, contain more time samples or variables than desired, are slow to access, and the user must manually combine ("aggregate") data that span many files (granules). Once these files are downloaded, format issues may further complicate using the data. Some data providers have developed servers that address some of these problems, but most do not scale to large datasets and require specialized software to be written to allow clients, machine or human, to easily use the data.

The reason for these difficulties is that scientific data collections developed by data providers have typically not been designed for analysis of, and access to, a single time series over a long time range.

The need for the TSDS server arose out of the domain of space physics, where time series are a very common data type used in analysis. In this field, the need for improved access to time series-like data is motivated by a frequent complaint from data users who state that they want

"all of the data over its full time range" if its size makes local storage reasonable, but most data providers only serve granules or subsets of the data [McPherron, 2008]. The need for TSDS is evident in many research papers where it is clear, based on where the data were obtained, that the researcher needed to do extensive aggregation manually using data from multiple data sources [Weigel and Baker, 2003; Vassiliadis et al., 2003; Green et al., 2004; Weimer, 2005; Bargatze et al., 2005; Tanskanen et al., 2005; O'Brien and Lemon, 2007; Rigler et al., 2007; Bortnik et al., 2007; Gjerloev and Barnes, 2008; Pulkkinen et al., 2008a; Pulkkinen et al., 2008b; Lemon et al., 2008; McPherron et al., 2009].

TSDS specializes in removing the need for the scientist to perform time consuming or complex pre-processing steps on time series-like data. The key improvements that it provides are (1) a decrease in the time required to fulfill a request, and (2) a simple API that can access data from many existing data collections and hides the implementation details of how the data are stored.

Although the Virtual Observatory network in heliophysics [Fox, 2008] is partitioned by science domain, we believe that services on data fall more naturally into categories given by the structure of the data and how it is typically visualized and analyzed. Nearly every science domain has a need for analysis of time series-like data, and most data providers develop ad-hoc solutions, presumably because software such as TSDS is not available.

The problem of optimally serving and filtering data is best solved based on the types of filtering and transformation operations that that are applied to the data, of which there are three general categories: time series (including spectrograms), 2-D (as in maps and contour plots), and 3-D (typical for the output from a simulation). In this work we deal only with time series-like data, independent of the science domain from which it originates. The motivation is that each of the three listed types have fundamentally different filtering operations that are typically applied, and there are fundamentally different storage and retrieval approaches for each. (More specific classifications have been proposed [http://www.unidata.ucar.edu/software/netcdf-java/CDM/CDMpoints.doc].) For example, 3-D data is typically viewed for only a few samples in time, slices through a volume are plotted, and the user has the option of rotating the view. In contrast, time series are typically viewed as power spectra, histograms, and stack plots, and the user interacts with the data by zooming in and out.

In accordance with the usage patterns for time series data, our server provides support for the following types of time series. (Technically there is overlap in the following categories, but in science discussions they are often referred to separately.)

- Scalar, vector, and matrix - each time step has one or more values associated with it. The values typically have an associated unit vector or set of coordinate labels; and
- Spectrum - each time stamp has a set of frequencies associated with it. Frequencies can be non-uniformly spaced and each [frequency, time] pair has an associated amplitude.

In this work we describe the design of TSDS, its implemented features, and a few of the planned features. This is an on-going project being developed in collaboration with the developers of OPeNDAP and software engineers working on data service issues in the heliophysics virtual observatory community.

# 3. Uses and Use Cases

TSDS allows for the implementation of many services for time series data. In this section we highlight a few of the applications or services that could be enabled by a server using TSDS.

### Multi-scale data set browse

A traditional way of dealing with high-resolution time series data is to store it in daily files and then have the user view pre-generated images of the data. TSDS allows the user to browse 1-second-cadence time series that spans over 10 years by drilling down from a view that shows the full time range of available data. Instead of selecting a time range to view from a drop-down list or typing a date into a text box, the user simply draws a box around the time range of interest. We have developed an example application that implements a prototype server. The application was written for Autoplot [Faden et al.; this issue] and uses a number of features of the server to allow a 1-second-cadence time series that spans over 10 years (200 MB compressed data volume) to be interactively browsed. A demonstration implementation is given at http://autoplot.org/applet_demos.

### Data reduction

A certain subset of users, for example those who do climatological or statistical studies, do not always require full time resolution for their analysis. There are several options for providing them with data. One common approach is to develop a separate data product that is the averaged form of the data for select time window widths and with select statistics (e.g. standard deviation, median, etc.) that were pre-computed in each window. The limitation of this is that the user is restricted to the statistics that were pre-selected by the data provider. If they want a statistic that is not available in the list, they could (a) download all of the data and do the calculation themselves or (b) ask the data provider to reprocess all of the data files. TSDS allows the user to select the time averaging window as an option in the data request. The interface allows the user to select an averaging window of, for example, 10 seconds, 1-minute, 1-hour, or 1-day. In this case what will be returned is the averaged data along with the number of data points that went into the average. Besides averages, the user has the option of requesting the maximum and minimum values of the time series in a given time window. Another option is to select strides of data; for example they could request that only every 60th data point is returned from a 1-second-cadence data set.

### Data filtering

A common science use case is to search for periodicities in data. At present, doing this usually requires downloading all of the relevant data granules to the scientist's local disk, extracting the data, and then computing the power spectrum. TSDS implements a number of filters that may be applied to the data prior to being sent to the user. The most basic filter is the minimum and maximum values in a time window, and moving average filters. Planned filters include the raw power spectrum (periodogram) and additional smoothing filters. A repository of filters will be made public along with instructions for submitting user-contributed filters to be made available through the service.

### Collaborative data cleaning

A common step that is required for the use of any data is to clean or de-spike it. The TSDS source code includes a simple template that allows a user to specify a data cleaning algorithm that can be added easily to the list of available filters.

### Data constrainer

TSDS includes an option that allows the user to specify "constraints" whereby they can request data when time series *A* was less than a given value and time series *B* was greater than a certain value.

### Index creator

Using TSDS it would be possible to easily create simple "derived" data sets comprised of combinations of time series. For example, if a scientist wanted to create a crude geomagnetic index, they could do this by sending a request to a TSDS server that specified a list of time series that should be added and that only the sum should be returned.

## 4. Implemented Features

### Filters and Other Operations

The TSDS API, which is described in section 8, allows a filter to be applied to the requested time series. At present, the following filters are available:

- Stride extraction - return every Nth time value;
- Thin - apply a stride such that approximately N time samples are returned;
- Replace - replace any occurrence of one value with another;
- Exclude - exclude any time sample that has a missing value; and
- Inequality (or relational) filters - return data when time and/or amplitude satisfied an inequality, e.g., (value>10.0 and time>2000-01-01T00:00:00.000) and (value>10.0 and time<2000-01-02T23:59.59.999)

In general, algorithms that require only one time sample for each operation are straightforward to implement, and the TSDS source code includes samples for creating such filters. Other planned filters are discussed in Section 5.

### Data Accessors

A data provider that seeks to provide a single API to a large set of existing time series data collections will require many data accessors (code that accesses data from a data collection and translates it into a common internal representation). TSDS reads the data that it serves into the Common Data Model (CDM) using the NcML (NetCDF Markup Language) conventions and associated NetCDF software libraries [http://www.unidata.ucar.edu/software/netcdf/ncml/]. NcML is an XML description of a dataset and can be thought of as a recipe that describes how to create a dataset, where the ingredients are URIs to data locations, and the instructions for each ingredient is a description of how to translate data from each URI into the CDM. When a dataset described in NcML is requested, the NetCDF software reads the dataset's NcML file and delegates to the appropriate Input/Output Service Provider (IOSP) required to read data from each URI into the CDM. To create an IOSP, one needs to specify how to relate metadata about variables and dimensions in a data granule from a data collection to the CDM and how to read a subset of a given variable into the CDM.

We have developed a number of IOSPs for TSDS. They read and write data to and from (1) specially formatted ASCII tables, (2) the fast database (TSDB) described in section 7, and (3) SQL databases. In a near-future release, we will interface to the API of Autoplot [http://autoplot.org; Faden et al., this issue], which connects to data stored in it's CDM equivalent, the QDataSet. The QDataSet representation of a time series is similar in scope to that of the Common Data Model. Autoplot's data accessors can aggregate time series data stored in files and directories that (1) follow common naming conventions based on the time coverage of the data it contains and (2) common file structuring conventions. The file formats it can access include CDF (Common Data Format; [http://cdf.gsfc.nasa.gov/]), NetCDF, CEF (Cluster Exchange Format; [http://www.space-plasma.qmw.ac.uk/csds/]), many ASCII table formats, binary table files, DataShop [http://datashop.jhuapl.edu/], and others.

## Fast database

In order to enable the science use-cases described in section 3., the response must be fast. Typical feedback from users on existing web-based interfaces that provide value-added services to time series data is that they are often too slow to be of use for interactive browsing [McPherron, 2008]; the users end up downloading the data locally, writing their own custom accessor, and developing their own browse, filter, and analysis tools. The most common approach for browsing heliophysics time series in a web browser is to page through plots that contain one day of data. This approach is presumably provided because the data collection contains data granules with one day of data per file, and the software required for providing a visualization is straightforward to implement.

There are two general approaches to providing uniform access and value-added filtering to data that span long time ranges:

1. "run-on-demand": A reader (or "accessor") for each data provider's data collection is developed that downloads data to the user's computer, extracts the relevant parts, and puts the data in an array or structure in the memory of the user's local machine; and
2. "pre-caching": The data are stored in a uniform manner on an intermediate server.

The "run-on-demand" approach has the weaknesses that the time to fulfill a request requiring data from multiple data providers will be limited by the slowest server, accessing data from multiple data sources increases the probability of a failure, and requests for similar data products will result in redundant server and bandwidth usage (if many users make similar requests that require a significant amount of CPU and disk usage by the server). A strength of this approach is that it does not require the additional storage of a copy of the data.

The "pre-caching" approach has the disadvantage that additional disk space is required. To store data from multiple data providers in a manner that allows fast response times requires duplication of the original data. Another potentially significant issue involves synchronization of the cache when there are changes to the original data collection. A data provider that uses TSDS to serve data from various data collections will have several options. First, they can re-run the caching procedure and write over the previous data. In this case the user who requests data with the REST API described in section 8 may receive different data on different days. Alternatively, the data provider could implement their own versioning system that interfaces with TSDS; we have identified this as a high-priority feature for future development and inclusion in the TSDS code base. The proposed versioning feature is described in section 5. Our suggestion is that the data provider allows access to old versions of time series but sets up their web interface to a TSDS so that the default is for the user to download the most recent version of the data. Keeping previous versions of time series could create data volume problems, but in section 6 we note that there are potential solutions, which we are also considering for future implemention.

The "pre-caching" approach has many advantages. As discussed in this section, and based on our experience, response times can be increased by up to a factor of 10-1000 for certain data sets (examples are given in section 6), the user is less likely to experience a failure in their request because only a single access point is involved, problems with corrupt data files are taken care of before the user makes a request, and when data are stored in a uniform manner that allows fast access, new filtering and data mining operations become possible.

The "pre-caching" and "run-on-demand" approaches are not mutually exclusive. A "pre-caching" approach requires a "run-on-demand" service to build the cache. Our claim is that for a significant amount of time series data, one can perform "run-on-demand" requests far in advance in order to allow fast and reliable responses and to enable compelling and responsive user interfaces for time series data.

When explaining the pre-caching approach, some have noted that it appears that a data center is being created because the cache is typically centralized; such centralization is not consistent with the "small-box" VO approach [http://hpde.gsfc.nasa.gov/VO_Framework_7_Jan_05.html] in which requests for data products are made from a VO interface and a list of URLs to data files is returned. TSDS should not be thought of as a data center creator, but rather a tool that can be used to provide and enable high-level services on time series data from a data server.

The following points provide other technical motivations for using the pre-caching approach:

1. When a data collection is created by a data provider, it is not typically optimized for speed. To gain speed, each data provider can optimize how their data collection is stored or accessed, but this usually requires additional storage and/or a significant amount of additional software development. Our approach is to optimize for speed on data for only one data collection, the cached data collection.

2. Consider the user who wants to look at one variable that is at a uniform 1-second resolution over a time span of 1 year. The data set spans 10 years. Most heliophysics data collections would have this variable stored in a file with 20 other variables, with one day of data per file. In order to serve this request, the server would need to visit 365 files, extract one variable and then push the data out. This takes a significant amount of CPU and requires many disk reads, which are very slow. This works for studies of an event or two that spanned a day or two, but not for long-term climatological studies. If one samples existing interfaces to time series data, they will invariably see some limit on the time range that can be requested. This constraint is almost always a result of the fact that accessing all parts of the time series requires significant CPU and disk access even if the data volume that is returned to the user is on the order of 100s of MB.

In section 6, we outline the implementation of the database format of the cache. This cache is referred to as the Time Series Data Base (TSDB).

# 5. Planned Features

In this section, we highlight several key features or services that are missing from the list of existing features that should either be used or supported by TSDS. In addition to describing the feature or service, we also provide suggestions for possible 3rd-party software that may provide some or all of the required functionality.

## Filters

- Time re-sampling - This filter will allow a time series to be re-sampled to a higher-resolution time grid. This will allow the user to have, for example, 64-second-averaged data re-sampled so that it is on the same time grid as 1-minute data;
- Time padding - This filter allows fill data to be returned when data is requested in a time interval that starts or ends before data exist for the time series;
- Averaging - A time averaging window is specified and all data are averaged in a sliding window with adjustable slide steps. NaN or fill values are omitted from the average; and
- Counting - For each window of a specified length, the number of values that satisfy a certain criteria is returned. One use case of this is to determine the number of valid data points that went into an average.

## Versioning

The typical versioning approach in heliophysics data management makes uniquely identifying a time series given only a time range and parameter name difficult. Two typical approaches are:

1. Data are stored in a file which has a fixed name, but its contents may change. When the contents change, the old version is not typically made as readily available as the most current version (e.g., the data are moved offline or moved to a different location); and

2. Data are stored in files, each with a version number in the file name. Each file typically contains one day of data. If data values in a few of the files change, the version number in the filename of only the affected days change. The old data files may or may not be made available at the same URL that they were originally accessed. For a user that studies long time ranges of data, the only way of uniquely specifying a data set is to list the version numbers of all of the files used. This becomes cumbersome when many years of data are analyzed.

The motivation for these versioning approaches is that the required storage space is small. If only one element of the time series is changed, only one file needs to be created or modified. A time series-based versioning approach makes more sense from the science perspective, as scientists think in terms of parameter names and not a list of files required to create a parameter.

Our planned approach is to version time series instead of files. In this way, the user will be able to uniquely specify a time series with a single ID, as in

`tsds://DataProviderName/DataSetName/TimeSeriesNumber/Version/StopDate`

where the StopDate entity is used to indicate the end date of the segment of time series being discussed. Because start dates do not typically change, they are not indicated in the ID. If the start date changes, the version number will be incremented.

In the case of the caching database, each ID corresponds to a physical file with a name

`DataProviderName_DataSetName_TimeSeriesNumber-v#.bin`

where "#" is a non-zero-padded integer and is the same as the number that appears in the "Version" slot of the TSDS ID.

Using this versioning approach, a new file is stored for each version of a time series, which may require a significant amount of extra storage space. For example, if a new version of a time series is created in which only a few values change, two files would need to be stored that have a great deal of overlap in content. In section 6, we outline a possible solution to this problem.

This versioning scheme does not specify the relationship between the data collection administrator and the entity that serves its data from a TSDB through TSDS. Ideally, the two provide updates to each other when changes are made to their respective holdings.

## Logging and Version Mappping

The TSDS versioning scheme is such that a given time series will have a version number that maps to a list of input data files, which typically have their own version numbering scheme. Certain users may want the ability to determine the specific list of input files that were used to generate a given TSDS time series version. To do this, an interface is needed that takes the inputs of a time series name and version and returns a list of files or API calls that were used by TSDS to create the full time series.   In addition, upon constructing a time series, TSDS would need to log the list of files or API calls that were used to create a given time series.

## Validation

Each version of a time series in the TSDB has an associated MD5 (Message Digest algorithm 5) checksum that is stored in the NcML metadata described in the next section.  The checksum is a 128-bit value that is a (virtually) unique fingerprint for a file.  The checksum can be computed on the client-side and then verified by that calculated on the server after the parameter download has completed.

## User Submission

Users often note that although an interface to time series data provides a comprehensive set filtering operations, it may be of limited use to them because they need to apply this filter to data that is not available through the interface.   In this case the user will revert to their old manual data access approach in which they write a data accessor to access the data and write their own local filtering software.  This problem can be addressed by allowing the user to submit their own data to be compared side-by-side with existing data.

The need for a submission service was motivated by our experience with users as a part of ViRBO [http://virbo.org] and participation in the Geospace Environment Modeling Radiation Belt Climatolgy Focus Group [http://virbo.org/GEMFG9].  Many users have derived data sets that they would like to expose to the community (most of which is composed of time series-like data).  Our initial suggestion was that they place all of their data in CDF [http://cdf.gsfc.nasa.gov/] files (CDF is a scientific data file format similar in scope to netCDF and HDF).  The feedback was that the amount of research required to do this was too large.  This is usually a barrier that is too high for the data-producing scientist [Crooker, 2006].  It requires learning about the "data model" of the file format.  To understand the data model the scientist must first learn the language and terminology of how data are described in the file format, which may include such concepts as strides, hypercubes, vary directions, and extendible variables.   As a result of this barrier, it is our experience that the typical approach for a scientist or data producer is to make files available in a documented format that is similar to that used by the project internally.

At minimum, the user submission service would allow a data producer who has their data in a data structure in their software analysis program on their local machine to submit their data with a simple two-step process:

1. Call a function or procedure called "`put_data`", as in `put_data(T,X,'A Data Set')`, which returns a URL to a web page with a form with metadata elements; and
2. Edit the metadata form and review a preview plot of the data

Once the metadata has been validated, the user may now tell another scientist that uses MATLAB, IDL, or Java that they may access this data using a command such as

```
3. [T,X] = get_data('A Data Set',[2001,1,1],[2003,1,12])
```

which makes a call to a server running TSDS. As with implementing a logging service for version mapping, this too is a complex task. There are many issues that must be addressed including provenance, permissions, upload limits, validation, and security. Initial research indicates that the RAMADDA [http://www.unidata.ucar.edu/software/ramadda] system could provide some of the desired user submission features.

## 6. TSDB Implementation

A key feature of TSDS is its interface to a fast data base for serving time series, called the Time Series Data Base (TSDB). (Not all data served through TSDS must be stored in a TSDB, however. The only requirement for serving data through TSDS is that an Input/Output service provider exists for the data set.)

Many time series server systems have substantial functionality for exploring, visualizing, or analyzing data, but have very high response times for large datasets. Requiring users to wait many seconds or even minutes for a response is a barrier to use. The motivation for TSDB is to overcome this barrier by providing highly responsive access to the data using a client-server method.

In this section we describe details about the layout of the caching database that enables the fast access, filtering, and response times. Standard relational databases that implement SQL are not designed to be fast for the type of operations that our service will provide, which has been confirmed by experimentation. Our caching database is file based, but it is important to note that these files are not intended to be accessed or manipulated directly by scientists; access to their contents is through an API.

The key design requirement is that the caching database should be optimal for accessing or sub-setting a long time series. Based on this requirement, each time series parameter should be stored as a single file. This may seem counterintuitive to scientists that are familiar with a single file containing a set of 10-50 scientifically related parameters. However, the fact that our internal database uses one parameter per file is not a concern to the scientist, because they will access the data in these files through an API instead of by directly downloading the file and writing an accessor for it.

When one time series (of all types, including spectrograms) is stored in a single file, reading the file requires a single file open and single disk seek operation. In contrast, the number of such operations is on the order of 100-1000 for aggregating the files in order to create time series using data in traditional data collections.

As a practical example of the time savings when a full time series is stored in a single binary file, for the use case that the user wants to view or analyze a large time series that spans a very long time range, consider the task of extracting the 1-minute-cadence magnetic field over a time span of 10 years for a ground magnetometer station using data from http://www.wdc.bgs.ac.uk/catalog/master.html. The data collection layout is such that one month of data for a given magnetometer station (usually three or four components of the magnetic field) is stored in a single ASCII-formatted file, and one year of data are stored in a zip archive. To extract the data for one component of the magnetic field vector over a 10-year time span and read it into an array in MATLAB or IDL takes approximately 20 seconds. In contrast, if the full time series is stored as a flat binary file, the data can be read into MATLAB or IDL in less than 0.007 seconds; if the data are stored compressed, the read time is 0.07 seconds (the flat binary format and compression format are described below). Of course, we could attempt to optimize the MATLAB and IDL extraction code to reduce the read time from 20 seconds, but we would still be limited by the fact that the initial data are stored as text and every request to read these data into memory would require a conversion from text to binary.

Even in the case where the initial file format is binary, it is our experience, for the problem of accessing large time series that spans a long time range, the speed-up is substantial. For example, a MATLAB program that loops over a collection of 365 CDF files and extracts a scalar time series on a 16-second time grid took 10 seconds to put the data in memory. In contrast, loading this variable as a flat binary array takes less than 0.002 seconds if the variable was stored on disk as a flat binary array and less than 0.02 seconds if stored with compression.

The point of these examples is that the structure of existing time series data collections is unsatisfactory for use cases that require quickly viewing or analyzing a large time series that spans a long time range. If multiple users want to view a 16-second time series over a full year, it makes sense to do the CPU and disk-intensive task of extracting the time series only once and then cache the result.

If the file associated with a time series is large, then one would naturally want the TSDB to have the capability of using compression. We have investigated a number of compression techniques and have concluded one of the easiest implementation options is to store the file on a compressed file system, because they allow random access to a compressed file (that is, the entire file does not need to be uncompressed in order to access a small chunk of data in the middle of the file). Based on testing on a few compressed file systems (squashfs and cloopfs), we have found that the time required to write a file is about ten times longer than that required to store the same data in an HDF5 file with compression enabled. Our interim, and possibly final, solution is to use the HDF5 file format to emulate a compressed file system with random access capabilities. That is, although below we define a .bin file format, large time series in a TSDB are actually stored in HDF5 files using compression and a request for TSDS to read data from a large .bin file is intercepted by a function that pulls the required data from a HDF5 file.

The time series server database (TSDB) .bin format uses one logical file per parameter, where a parameter is a scalar, vector, or spectrogram (with time-independent frequency components), containing a list of 64-bit IEEE 754 floating point values. A scalar time series is simply a binary file with N values, where N is the number of samples. The corresponding time stamps are not stored in the same file.

A vector measurement would be stored as

```
Vx(t=0),Vy(t=0),Vz(t=0), ..., Vx(t=N-1),Vy(t=N-1),Vz(t=N-1)
```

where `t` corresponds to the measurement time. A spectrogram where the amplitudes are always measured in the same four frequency bins would be stored as

`A(f1,t=0),A(f2,t=0),A(f3,t=0),A(f4,t=0),...,A(f1,t=N-1),A(f2,t=N-1),A(f3,`

If measurements for the time series were not made at a uniform cadence, there would be a file that contained the N timestamps

`t(0), ..., t(N-1)`

The internal structure was designed based on the assumption that long time spans of data are to be viewed or analyzed. These files are not intended for scientific exchange or download format, in part because they do not contain metadata. Associated with TSDB bin file is an NcML file that follows the conventions described in the next section.

Every time series should be viewed as a flat binary array with a uniform data type (always 64-bit IEEE 754 little endian). For example, temperature was recorded at the start of every hour for a day, its TSDB file would be 8*24 bytes. If 24 temperature measurements were taken at a random times during the day, two files would be needed: a time stamp file of 8*24 bytes and a temperature measurement file having the same size.

For data on a non-uniform time grid, the time variable is stored as 64-bit IEEE 754 floating point values and the connection of the time file to a data file is made using NcML. The metadata for the time file includes the start date of the first data point and the time unit. The file contains values that represent the time since the start date and in the time units specified in the NcML file.

In section 5, we have identified the importance of versioning. A feature that we plan on implementing involves reducing the space required to store previous versions of time series. The "best practices" convention for REST-style URLs that interface with a database is that a given URL should correspond to a unique response. For scientific purposes this is important because it is difficult to reproduce results if previously used versions of data are no longer available. There are several possibilities for minimizing the space required to store previous versions. One option is to allow TSDS to use the features of a versioning file system that stores incremental diffs, if available. A second option is to store the differences between the base version in an HDF5 file and use the "pipeline filter" feature of HDF5, which allows a mathematical transform to be applied to data before it is returned without significant memory or time overhead. In many cases, changes to versions in time series involve changes that result in a difference time series that is highly compressible, such as a slowly changing waveform or a constant value added to portions of the time series.

## 7. NcML and the TSDS conventions

Each time series stored in a TSDB has an associated metadata file that contains the minimal amount of "use" metadata that is required for TSDS to do a mathematically sensible computation on the numbers or for the numbers to be rendered on the screen as either a time series, vector, or spectrogram.   The minimal amount of information that is needed to do a mathematically sensible operation on a time series include time tags and the type of time series, e.g., scalar, vector, or spectrogram. The "use" information for each time series is stored in an NcML file [http://www.unidata.ucar.edu/software/netcdf/ncml/]; we have added several conventions to NcML to form the TSDS conventions for NcML.  These metadata records contain a URL to "science interpretation" metadata in the "ScienceMetaData" attribute.  The ScienceMetaData URL is intended to be a link to metadata formatted in a domain-specific schema such as SPASE [http://spase-group.org] or FGDC [http://www.fgdc.gov/]; if no such metadata is available, this link may be to a README file or a web page with information that may be useful to the scientist.

Because the TSDB metadata records contain only minimal "use" information, one or more filtering operations may be requested that do not make sense physically.   We have developed TSDS and TSDB to handle the non-science-specific aspects of time series data and the ScienceMetaData attribute is intended for use by higher-level software that makes use of data served through TSDS and places constraints on the types of requests that can be made.

An example of a minimal metadata record is as follows:

```
<netcdf>
    <attribute name="title" value="Variable1 (SourceAcronym Subset1 1-hour)"/>
1.  <attribute name="Conventions" value="COARDS,TSDS"/>
2.  <attribute name="TSDSID" value="tsds://DataProviderName/DataSetName/TimeSeriesNumber/Version/StopDate">
3.  <attribute name="ScienceMetaData" value="URL"/> <!--Optional-->
4.  <attribute name="DataType" value="time_series"/> <!-- time_series, vector, or spectrogram-->
5.  <attribute name="StartDate" value="1989-01-01"/>
6.  <attribute name="StopDate" value="2005-12-31"/>
7.  <attribute name="MD5" value ="...">
8.  <attribute name="PointsPerDay" value="24"/> <!--Optional-->

    <aggregation type="union">
        <netcdf>
            <dimension name="time" isUnlimited="true" length="149016"/>
            <variable name="time" shape="time" type="double">
                <attribute name="long_name" type="String" value="time"/>
                <attribute name="units" type="String" value="minutes since 1989-01-01 00:00:0.0"/>
                <values increment="1.0" start="0.5"/>
            </variable>
        </netcdf>

        <netcdf>
            <dimension name="data" isUnlimited="false" length="149016" />
            <variable name="Variable1" shape="time" type="double">
                <attribute name="long_name" type="String"
```

```
                    value="Variable1"/>
9.                  <attribute name="cformatstring" type="String" value="d"/> <!--Optional either single or comma separated list-->
                    <attribute name="units" type="String" value="Hour"/>
                    <values increment="1.0" start="0"/>
               </variable>
          </netcdf>

     </aggregation>
</netcdf>
```

This NcML file fully specifies a time series. The TSDS attributes are labeled 1-9. Line 3. contains the URL to metadata that could be used to interpret the data. The Start and Stop dates indicate the days that the first and last measurement were taken, respectively. The PointsPerDay is used to indicate the number of timestamps per day and may be an estimate for non-uniformly spaced data. The cformatstring is an optional parameter that is used by the server when rendering the data in an ASCII table. The intended use is for cases where the initial data were all represented in a certain format in ASCII; this specification allows the default ASCII output rendering to be the same. If only one character is given, it is applied to all values. Otherwise it may be a comma-separated list with one character per dimension in the corresponding vector or frequency array in the corresponding spectrogram.

The time series specified by the NcML above is simply the numbers 0, 1, 2, … on a uniform 1-minute time grid starting at 1989-01-01T00:00:30.00000. A more common case would be to replace the second `<netcdf>` block above with

```
<netcdf location="DataProviderName_DataSetName_TimeSeriesNumber-v#.bin" iosp="tsdb.iosp.BinIOSP">
     <dimension name="time" isUnlimited="true" length="149016"/>
     <variable name="Variable1" shape="time" type="double">
          <attribute name="long_name" type="String" value="Variable1"/>
          <attribute name="cformatstring" type="String" value=".16f"/>
          <attribute name="units" type="String" value="Unit"/>
          <attribute name="_FillValue" type="double" value="NaN"/>
     </variable>
</netcdf>
```

This block specifies that the data are to be found in the file data.bin. (The .bin format is described in the previous section.) The IOSP tag is used by NetCDF software to delegate the reading of data.bin. Because the time block specifies a uniform time grid, this is the only information that TSDS needs in order to serve or do a mathematical operation on the time series stored in data.bin. For data on a non-uniform grid, the time block would be replaced with a pointer to a file of time stamps such as

```
<netcdf location="DataProviderName_DataSetName_TimeSeriesNumber-v#.bin" iosp="tsdb.iosp.BinIOSP">
     <dimension name="time" isUnlimited="true" length="149016"/>
     <variable name="time" shape="time" type="double">
```

```
        <attribute name="long_name" type="String" value="time"/>
        <attribute name="units" type="String" value="minutes since 1989-01-01 00:00:0.0"/>
    </variable>
</netcdf>
```

# 8. API

In order to provide both fast response times and a flexible interface, there are three access modes for TSDS, numbered in terms of decreasing speed and increasing ease of use. Mode 1 is primarily intended for situations where speed is needed (and the data have already been stored or cached in a TSDB); in these modes data are directly accessed from the a TSDB and the requested data is not first re-formatted into the Common Data Model (CDM). Mode 2 provides high-level abstractions and high versatility but require TSDS to first transfer time series from a data collection in its native storage form into the CDM. In a future release, users will have the ability to store data in a TSDB by making the appropriate Mode 2 requests to a TSDS server.

Each mode serves data based on a RESTful HTTP request.

## *Mode 1*

Mode 1 is designed to provide the fastest access to data. In Mode 1 the server takes inputs of

- URL: The URL of a TSDB binary file
- StartIndex: The first value to access
- StopIndex: The end value to access (optional)

The output is simply (`StartIndex - EndIndex +1`) 64-bit little-endian IEEE-754 formatted values. The full physical TSDB file may also be accessed through the URL

```
http://host/tsdb/
DataProviderName_DataSetName_TimeSeriesNumber-v0.bin
```

Note that the version number in this example is zero, and all files in the TSDB have the string "-v#.bin", where "#" is a non-zero-padded integer. The "use" metadata (NcML with TSDS extensions) for this parameter is stored at

```
http://host/tsdb/
DataProviderName_DataSetName_TimeSeriesNumber-v0.ncml.
```

The syntax is to access the data is

```
http://host/tsds/
DataProviderName_DataSetName_TimeSeriesNumber-
v0.bin?[StartIndex:EndIndex]
```

which returns a TSDB binary-formatted stream. If the end-of-file is reached before the (StartIndex_file-EndIndex_file+1) values are returned, the server returns NaNs. An error is returned if StartIndex is less than zero or if EndIndex is less than StartIndex.

## *Mode 2*

Mode 2 provides the greatest degree of versitility and ease of use. In this mode users can specify a time range, select a variable, and apply filters to the resulting time series. This flexibility is achieved by first converting data associated with a request into the Common Data Model. This conversion imposes an additional processing cost, but this cost could be mitigated by storing data in a TSDB after the first time it is requested. (Note that only time series served from TSDS that originate from a TSDB have a unique TSDS ID.)

Mode 2 uses the TSDS OPeNDAP server which is implemented as a Java Servlet. It takes OPeNDAP-compliant URL requests [http://opendap.org/user/guide-html/guide_33.html] of the form:

```
http://host/tsds/dataset.suffix?constraint_expression
```

for data sets that are not in the TSDB (and DataSetName may be the name of an actual file on disk that does not follow the TSDB naming conventions); for time series that are served from the a TSDB, the syntax is

```
http://host/tsds/DataProviderName_DataSetName_TimeSeriesNumber-
v#.suffix?constraint_expression
```

where

- host: name of the computer hosting the TSDS servlet,
- dataset: name of a dataset to be served,
- suffix: type or format of the output, and
- constraint_expression: A collection of request parameters such as time range and filter.

and the suffixes (i.e., output options), include

- info: information about the dataset and request options,
- html: HTML view of dataset information and a form for requesting data,
- dds: dataset Descriptor Structure (ASCII),
- das: dataset Attribute Structure (ASCII),
- dods: data object as defined by the Data Access Protocol (DAP), and
- asc: data object represented as ASCII.

Other output options include

- csv: comma separated values,
- dat: tabular ASCII format,
- bin: TSDB binary,

- ncml: associated NcML file (if applicable or if data request was to a data set in the TSDB),
- nc: Network Common Data Form (NetCDF) file,
- h5: Hierarchical Data Format (HDF) version 5, and
- json: JavaScript Object Notation (JSON).

Constraint expressions consistent with the OPeNDAP specification are allowed, e.g,

`var1,var2&time>t1&filter()`

where

- var1,var2: a list of variables to return (default is all variables associated with the dataset)
- time>t1: a time range constraint where t1 is in native time format or ISO 8601 (YYYY-MM-DDThh:mm:ss)
- filter(): an algorithm to be applied to the data on the server before being sent to the client

By default, if an end date is not specified the last date of available data is assumed.

The current implementation only allows one filter to be applied, but a future release will support filter chaining. Note that the OPeNDAP protocol allows multiple operations and constraint expressions to be applied in a single URL [http://opendap.org/user/guide-html/constraint.html]. The following are some of the URLs that are in use by a data provider using an early version of TSDS [http://lasp.colorado.edu/LISIRD/]. (Note that these data do not originate from a TSDB, so the dataset naming convention of `DataProviderName_DataSetName_TimeSeriesNumber-v#` is not used.)

- `http://host/tsds/TSI.csv?time>2003-02-25&time<2009-03-27`
- `http://host/tsds/sunspot_number.csv?&replace_missing(NaN)&time>2003-02-25&time<2009-06-01`
- `http://host/tsds/TSIdb.csv?time,correctedIrradiance&replace_missing(NaN)&time>2007-07-11&time<2008`
- `http://host/tsds/f107.csv?&replace_missing(NaN)&time>2005-08-16&time<2005-10-05`
- `http://host/tsds/sorce_ssi.csv?time,irradiance&time>=2009-01-01&time<2009-01-02`
- `http://host/tsds/sorce_ssi.csv?wavelength,irradiance&time>=2009-01-01&time<2009-01-02`

## 9. Server-Side Software

Most of TSDS is implemented using Java Servlet technology, although Perl/CGI is used for Mode 1. Installation involves placing a single .war file into any Java servlet container, such as Tomcat, and adding several options to an Apache server configuration file.

## 10. Software Files, Availability, and Requirements

All required software files are available at http://sourceforge.net/projects/tsds/ under the GNU Public License, version 2.0. TSDS requires a J2EE Servlet container (such as Apache Tomcat). At the time of publication of this paper, version 1.0 of the software is available in source code form. Capabilities (and to some extent APIs) are expected to evolve rapidly over the next year after which a version 2.0 will be released.

At the time of publication, a demo server of TSDS will be available for testing at http://timeseries.org/. The initial list of data that will be made available there will include data from the Virtual Radiation Belt Observatory [http://virbo.org/] and select data sets from the Interactive Solar Irradiance data center at the Laboratory for Atmospheric and Space Physics [http://lasp.colorado.edu/LISIRD/].

## 11. Acknowledgments

Development was supported in part by NASA grant NNX07AB70G (VxO for S3C Data: The Virtual Radiation Belt Observatory) and NSF grant 0457577 (FDSS: Faculty Development in Space Weather Research: A Systems Perspective).